\documentclass[twocolumn,english]{revtex4-1}
\usepackage[T1]{fontenc}
\usepackage[latin9]{inputenc}
\usepackage{color}
\usepackage{babel}
\usepackage{bm}
\usepackage{amsmath}
\usepackage{amssymb}
\usepackage{graphicx}
\usepackage[unicode=true,pdfusetitle,
 bookmarks=false,
 breaklinks=false,pdfborder={0 0 0},pdfborderstyle={},backref=false,colorlinks=true]
 {hyperref}

\makeatletter
\usepackage{babel}

\makeatother

\begin{document}
\title{Squeezing-induced Topological Gap Opening on Bosonic Bogoliubov Excitations}
\author{Liang-Liang Wan$^{1,2}$, Zixian Zhou$^1$, and Zhi-Fang Xu$^1$}
\email{xuzf@sustech.edu.cn}

\affiliation{$^1$Shenzhen Institute for Quantum Science and Engineering and Department
	of Physics, Southern University of Science and Technology, Shenzhen
	518055, China ~\\
	$^2$School of Physics and Technology, Wuhan University, Wuhan, Hubei 430072,
	China}
\begin{abstract}
We investigate the role of squeezing interaction in inducing topological
Bogoliubov excitations of a bosonic system. We introduce a squeezing
transformation which is capable of reducing the corresponding Bogoliubov-de
Gennes Hamiltonian to an effective non-interacting one with the spectra
and topology unchanged. In the weak interaction limit, we apply the
perturbation theory to investigate the squeezing-induced topological
gap opening on bosonic Bogoliubov excitations and find that the squeezing
interaction plays an equivalent role as a spin-orbit or Zeeman-like
coupling in the effective Hamiltonian. We thus apply this formalism
to two existed models for providing deeper understandings of their
topological structures. We also construct minimal models based on
the elegant Clifford algebra for realizing bosonic topological Bogoliubov
excitations. Our construction is potentially applicable for experiments
in bosonic systems.
\end{abstract}
\maketitle

\section{Introduction\label{sec:Intro}}

In this century tremendous progress has been made in the exploration
and realization of topological states of matter \citep{Hasan2010RevModPhys,Qi2011RevModPhys,Chiu2016RevModPhys}.
Besides the famous topological insulators and superconductors \citep{Bernevig2013,Sato2017RPP},
topological bosonic systems without fermionic analogy also attract
wide attentions \citep{Karzig2015PhysRevX,Yang2015PhysRevLett,Jin2016NatComm,Ozawa2019RevModPhys,Zhang2018AdvPhys}.
It is reported that magnetic materials \citep{Zhang2013PRB,Shindou2013PRB0,Shindou2013PRB,Shindou2014PRB,Chisnell2015PRL,Li2017PRL,Wang2017PRB,Nakata2017PRB,McClarty2017NatPhys,McClarty2018PRB,Lu2018arXiv,Kondo2019PysRevB,Kondo2019PRB2,Diaz2019PRL},
optical systems \citep{Bardyn2016,Peano2016NatCommun,Engelhardt2016PRL},
and ultracold atoms \citep{Engelhardt2015PhysRevA,Furukawa2015NJP,Xu2016PhysRevLett,Di_Liberto2016PRL,Luo2018PRA}
are capable of supporting topological bosonic Bogoliubov excitations
(BEs) with nontrivial Chern numbers or $\mathbb{Z}_{2}$ invariants.
These systems are described by a bosonic Bogoliubov-de Gennes (BdG)
model which is adaptive to the peculiar symplectic/pseudo-unitary
diagonalization \citep{Ring1980,Blaizot1986,Colpa1986PA}. Recent
works further show that their topological classification is different
from the common Atland-Zirnbauer (AZ) class \citep{Altland1997PhysRevB,Kitaev2009AIP,Stone2011JPA,Lu2018arXiv,DeNittis2019,Lein2019PRB,Zhou2019arXiv}.
Moreover, the edge state of the bosonic BEs as a typical topological
effect is also predicted \citep{Zhang2013PRB,Shindou2013PRB0,Shindou2013PRB,Shindou2014PRB,Engelhardt2015PhysRevA,Chisnell2015PRL,Furukawa2015NJP,Di_Liberto2016PRL,Bardyn2016,Peano2016NatCommun,Xu2016PhysRevLett,Engelhardt2016PRL,McClarty2017NatPhys,Diaz2019PRL,Malki2019PRB}.

The bosonic BdG Hamiltonian is a mean-field description of a bosonic
system with a weak nonlinear interaction \citep{Ring1980,Blaizot1986}.
It consists of a single-particle dispersion term and a bi-particle
annihilation/creation interaction known as the two-mode squeezing
\citep{Scully1997}. As a vast difference from the fermionic case,
the Bogoliubov transformation preserving the bosonic commutation relation
belongs to a non-compact pseudo-unitary group \citep{Simon1999JMP}.
This brings difficulty in the analysis of the BE modes and spectra.
For this reason, previous studies usually resort to numerical computations
and are unable to give a clear understanding on thetopological origin
of BEs with squeezing interaction. Besides, these
sporadic examples \citep{Zhang2013PRB,Shindou2013PRB0,Shindou2013PRB,Shindou2014PRB,Chisnell2015PRL,Li2017PRL,Nakata2017PRB,McClarty2017NatPhys,Wang2017PRB,McClarty2018PRB,Lu2018arXiv,Kondo2019PysRevB,Kondo2019PRB2,Diaz2019PRL,Bardyn2016,Peano2016NatCommun,Engelhardt2015PhysRevA,Furukawa2015NJP,Engelhardt2016PRL,Di_Liberto2016PRL,Xu2016PhysRevLett,Luo2018PRA}
do not provide a systematical way to construct such topological BEs.
The essential problem here is the missing of an intuitive picture
that depicts the role of squeezing in the BE spectra.

In this article, we look for a concise formula that characterizes
the topological BEs of bosonic BdG systems in the weak interaction
limit. Our exploration starts with a fact that the bosonic BdG Hamiltonian
can be continuously deformed onto a non-interacting one with the same
BE spectra and topology. We reveal that the non-interacting Hamiltonian
is nothing but the squeezed state representation of the original model
and acts as an effective Hamiltonian. In its explicit expression,
the perturbative squeezing interaction can be linearized as an effective
spin-orbit coupling (SOC) added to the dispersion term. This result
translates the topological analysis and construction of bosonic BdG
model to those of single-particle one belonging to the AZ class \citep{Kitaev2009AIP,Stone2011JPA}.

Based on this discovery, we notice that the present topological bosonic
BEs are of a unified construction. In those models the dispersion
part usually corresponds to a gapless band structure while the presence
of squeezing interaction opens a bulk gap in the high-lying BE spectra.
The construction has been applied to many topological band models
and can be viewed as a variation of the band-inversion \citep{Kane2005PhysRevLett,Kane2005PhysRevLett2,Bernevig2006PRL,Fu2007PRL}
phenomenon. In this situation, we find that the squeezing-induced
SOC in the effective Hamiltonian can be further approximated as a
momentum-independent Zeeman-like coupling. If the model is made up
of Dirac matrices, it can even be reduced to a constant mass term.
Therefore, the topological effect of squeezing has a clear fermionic
analogy and interpretation. This provides us a deeper understanding
toward the squeezing-induced topological gap on bosonic BEs.

As another motivation of this work, we also search for the minimal
construction of topological bosonic BEs. Like the Atiyah-Bott-Shapiro
construction for free-fermion systems \citep{Atiyah1964Top,Stone2011JPA},
here we utilize the Clifford algebra to yield the bosonic BdG Hamiltonian
including both the dispersion and squeezing terms. A squeezing-induced
gap is also designed to generate topological BE bands. Despite of
the difficulty in solving the exact BE spectra, our effective Hamiltonian
takes an elegant Dirac form and faithfully reflects the origin of
topology. This minimal construction is capable of systematically producing
topological bosonic BEs by converting the fermionic topological insulator
into the bosonic version. The method will also inspire more realistic
models for experiments in the future.

The rest of article is organized as follows. In Sec. \ref{sec:Model},
the bosonic BdG model and its effective Hamiltonian are introduced.
In Sec. \ref{sec:Pertur}, perturbation formula of squeezing-induced
topological gap is derived and applied to discussing two existed models.
In Sec. \ref{sec:Constr}, the minimal construction of topological
bosonic BE is built and two nontrivial models are designed. In Sec.
\ref{sec:Conclu} conclusions are made.

\section{Model and effective hamiltonian\label{sec:Model}}

We consider the bosonic BdG model which describes a bosonic system
with a nonlinear interaction after the mean-field approximation \citep{Ring1980,Blaizot1986,Rossignoli2005PhysRevA}.
This model is available in various physical systems including magnonic
crystal \citep{Shindou2013PRB0,Shindou2013PRB,Chisnell2015PRL,Kondo2019PysRevB,Kondo2019PRB2},
photonic crystal \citep{Notomi2008NatPhotonics,Eggleton2011NatPhotonics,Peano2016NatCommun},
and ultracold atoms in optical lattice \citep{Aidelsburger2013PhysRevLett,Aidelsburger2014NatPhys,Engelhardt2015PhysRevA,Furukawa2015NJP,Di_Liberto2016PRL,Xu2016PhysRevLett,Luo2018PRA,Yan2018PhysRevLett}.
The BdG Hamiltonian contains a dispersion term $\sum_{\boldsymbol{k}}\hat{\boldsymbol{a}}_{\boldsymbol{k}}^{\dagger}H_{a}\left(\boldsymbol{k}\right)\hat{\boldsymbol{a}}_{\boldsymbol{k}}$
that originates from the single-particle system and a part of interaction
\citep{Di_Liberto2016PRL,Xu2016PhysRevLett,Luo2018PRA}. It also contains
a squeezing interaction $\frac{1}{2}\sum_{\boldsymbol{k}}\hat{\boldsymbol{a}}_{\boldsymbol{k}}^{\dagger}H_{s}\left(\boldsymbol{k}\right)\hat{\boldsymbol{a}}_{-\boldsymbol{k}}^{\dagger\top}+\text{h.c.}$
that is reduced from the rest part of interaction. Here $\hat{\bm{a}}_{\bm{k}}^{\dagger}=(\hat{a}_{\bm{k},1}^{\dagger},\cdots,\hat{a}_{\bm{k},N}^{\dagger})$
is a multi-component bosonic creation operator with regard to $d$-dimensional
momentum $\boldsymbol{k}$, and $\hat{\boldsymbol{a}}_{\boldsymbol{k}}^{\dagger\top}$
transposes these elements (operators $\hat{a}_{\boldsymbol{k},i}^{\dagger}$)
to a column vector. The complete BdG Hamiltonian can be concisely
written as
\begin{equation}
\hat{H}=\frac{1}{2}\sum_{\bm{k}}\hat{\phi}_{\bm{k}}^{\dagger}H\left(\bm{k}\right)\hat{\phi}_{\bm{k}},
\end{equation}
with field operator $\hat{\phi}_{\bm{k}}^{\dagger}=(\hat{\bm{a}}_{\bm{k}}^{\dagger},\hat{\bm{a}}_{-\bm{k}}^{\top})$
and $2N\times2N$ Hermitian matrix
\begin{equation}
H\left(\boldsymbol{k}\right)=\left(\begin{array}{cc}
H_{a}\left(\boldsymbol{k}\right) & H_{s}\left(\boldsymbol{k}\right)\\
H_{s}^{\ast}\left(-\boldsymbol{k}\right) & H_{a}^{\ast}\left(-\boldsymbol{k}\right)
\end{array}\right).
\end{equation}
In this system, there is an intrinsic particle-hole symmetry \citep{Shindou2013PRB,Peano2016NatCommun,Lieu2018PRB,Zhou2019arXiv,Lein2019PRB}
which reads
\begin{equation}
U_{C}^{-1}H^{\ast}\left(-\boldsymbol{k}\right)U_{C}=H\left(\boldsymbol{k}\right),\,U_{C}=\left(\begin{array}{cc}
 & I_{N}\\
I_{N}
\end{array}\right),\label{eq:symm}
\end{equation}
where $I_{N}$ denotes the $N\times N$ identity matrix. Besides,
we require the energy non-negative (i.e., $H\left(\bm{k}\right)$
positive semi-definite) to guarantee the thermodynamic stability.

Before discussing the topology of BEs, we first need to solve the
BEs by a Bogoliubov transformation$V$ that preserves
the bosonic commutation relation
\begin{equation}
\left[\hat{\phi}_{\boldsymbol{k},i},\hat{\phi}_{\boldsymbol{k},j}^{\dagger}\right]=\tau_{ij},\,\tau=\left(\begin{array}{cc}
I_{N}\\
 & -I_{N}
\end{array}\right),\label{eq:Commutation}
\end{equation}
i.e., $V^{\dagger}\tau V=\tau$. It has been known
that $V$ forms a non-compact pseudo-unitary group ${\rm U}\left(N,N\right)$.
This has a vast difference from the unitary group used to diagonalize
a fermionic system. Nevertheless, the traditional approach leads
to implicit solutions and cannot reveal the effect of squeezing plainly.
Instead, we adopt a squeezing transformation which is capable of explicitly
transforming the BdG Hamiltonian to a non-interacting one \citep{Scully1997}.
The effective non-interacting Hamiltonian will not only give the BE
modes and spectra via a common unitary diagonalization but also present
the effect of squeezing in an intuitive way.

To be more specific, we define a squeeze operator 
\begin{equation}
\hat{S}=\exp\sum_{\boldsymbol{k}}\frac{1}{2}\left[\boldsymbol{a}_{\boldsymbol{k}}^{\dagger}w\left(\boldsymbol{k}\right)\boldsymbol{a}_{-\boldsymbol{k}}^{\dagger\top}-\text{h.c.}\right],
\end{equation}
with a natural constraint $w^{\top}\left(\boldsymbol{k}\right)=w\left(-\boldsymbol{k}\right)$.
By using commutation relation 
\begin{equation}
\left[\hat{\phi}_{\boldsymbol{k},i},\ln\hat{S}\right]=\left[W\left(\boldsymbol{k}\right)\hat{\phi}_{\boldsymbol{k}}\right]_{i},\,W=\left(\begin{array}{cc}
 & w\\
w^{\dagger}
\end{array}\right),
\end{equation}
we introduce a unitary transformation
\begin{equation}
\hat{S}^{\dagger}\hat{\phi}_{\boldsymbol{k}}\hat{S}=e^{W\left(\boldsymbol{k}\right)}\hat{\phi}_{\boldsymbol{k}}.
\end{equation}
The field operator after transformation is a linear superposition
of $\hat{\boldsymbol{a}}_{\boldsymbol{k}}$ and $\hat{\boldsymbol{a}}_{-\boldsymbol{k}}^{\dagger}$,
which still keeps the commutation relation (\ref{eq:Commutation}).
This implies that the squeezing transformation is a special Bogoliubov
transformation. Then the Hamiltonian in the squeezed-state representation
is given by 
\begin{equation}
\hat{S}^{\dagger}\hat{H}\hat{S}=\frac{1}{2}\sum_{\bm{k}}\hat{\phi}_{\bm{k}}^{\dagger}h\left(\bm{k}\right)\hat{\phi}_{\bm{k}},
\end{equation}
where 
\begin{equation}
h\left(\boldsymbol{k}\right)=e^{W\left(\boldsymbol{k}\right)}H\left(\boldsymbol{k}\right)e^{W\left(\boldsymbol{k}\right)}.\label{eq:hamil}
\end{equation}
To some extent, $h\left(\bm{k}\right)$ and $H\left(\bm{k}\right)$
are physically equivalent since they are connected via a congruent
transformation.

By choosing a special $W\left(\boldsymbol{k}\right)$ we can make
$h\left(\boldsymbol{k}\right)$ block-diagonal, that is $\tau h\tau=h$.
Using the identity $\tau W\tau=-W$, we can infer that the block-diagonal
condition is equivalent to $\tau H\tau=\exp\left(2W\right)H\exp\left(2W\right)$
whose solution is given by \citep{Zhou2019arXiv}
\begin{equation}
e^{2W}=H^{-\frac{1}{2}}\left(H^{\frac{1}{2}}\tau H\tau H^{\frac{1}{2}}\right)^{\frac{1}{2}}H^{-\frac{1}{2}}.\label{eq:W}
\end{equation}
Since $H$ is positive definite (thermodynamic stability),
$H^{\pm1/2}$ are well defined. Also, $H^{1/2}\tau H\tau H^{1/2}=(H^{1/2}\tau H^{1/2})^{2}$
is positive definite, then its square root is well defined, too. If
$H$ encounters a zero energy, one may utilize $H+0^{+}\cdot I$ instead
of $H$ and solve the limit of $h$.

Combining Eqs. (\ref{eq:symm}), (\ref{eq:hamil}) and (\ref{eq:W}),
we find that the particle-hole symmetry is inherited by $h\left(\boldsymbol{k}\right)$,
i.e., 
\begin{equation}
h\left(\boldsymbol{k}\right)=U_{C}^{-1}h^{\ast}\left(-\boldsymbol{k}\right)U_{C}=\left(\begin{array}{cc}
h_{a}\left(\bm{k}\right)\\
 & h_{a}^{\ast}\left(-\bm{k}\right)
\end{array}\right).
\end{equation}
We emphasize that the above treatment is equivalent
to that of Ref. \citep{Zhou2019arXiv}. But the language of second
quantization applied here gives a clear physical interpretation,
the BdG Hamiltonian in the squeezed state representation is reduced
to an effective non-interacting one 
\begin{equation}
\hat{S}^{\dagger}\hat{H}\hat{S}=\sum_{\boldsymbol{k}}\hat{\boldsymbol{a}}_{\boldsymbol{k}}^{\dagger}h_{a}\left(\boldsymbol{k}\right)\hat{\boldsymbol{a}}_{\boldsymbol{k}}.\label{eq:nint}
\end{equation}
Now we can easily achieve the BE modes and spectra of $\hat{H}$ by
the unitary diagonalization of $h_{a}\left(\boldsymbol{k}\right)$.
The BE modes are the eigenstates of the effective non-interacting
system after a squeezed-state transformation, and the excitation spectra
are simply the eigenvalues of $h_{a}\left(\boldsymbol{k}\right)$.

The squeezing transformation can also be applied to a more generic
quadratic bosonic Hamiltonian without the particle-hole symmetry \citep{Zhou2019arXiv}.
The obtained effective non-interacting Hamiltonian would contain two
isolated parts. Details are given in Appendix \ref{sec:Gen}.

The topological relation between the original and effective Hamiltonians
has been already revealed in Ref. \citep{Zhou2019arXiv}. Since the
squeezing transformation $\hat{S}$ can always be deformed into an
identity operator in a trivial way, $H\left(\boldsymbol{k}\right)$
can be continuously deformed onto $h\left(\boldsymbol{k}\right)$
while keeping the excitation gap opened and the symmetry invariant.
This means that they are homotopy equivalent. As a result, the topological
structure of $\hat{H}$ is identical to that of gapped non-interacting
system $h_{a}\left(\boldsymbol{k}\right)$, characterized by the AZ
class \citep{Kitaev2009AIP,Stone2011JPA,Chiu2016RevModPhys}. Therefore,
we only focus on $h_{a}\left(\boldsymbol{k}\right)$ instead of $\hat{H}$
in the following discussions. Then the topological analysis of bosonic
BEs is converted to that of fermionic topological insulator.

\section{Perturbation formulation\label{sec:Pertur}}

In this section, we aim to clarify the key role of the squeezing interaction
in inducing the topological gap opening on high-lying BEs by using
perturbation theory. As shown in Sec. \ref{sec:Model}, the effect
of squeezing has been encoded in the expression of effective Hamiltonian
$h_{a}\left(\boldsymbol{k}\right)$. Despite of the complicated form,
linearization of $H_{s}\left(\bm{k}\right)$ can be made in the squeezing
transformation when the squeezing interaction is weak. As a result,
a more concise formula of $h_{a}\left(\boldsymbol{k}\right)$ will
be achieved after perturbation truncation and a clearer physical picture
will be disclosed.

Firstly, Eq. (\ref{eq:hamil}) can be expanded with regard to $W$
and expressed as
\begin{equation}
h=H+\left\{ W,H\right\} +\frac{1}{2}\left\{ W,\left\{ W,H\right\} \right\} +\cdots.
\end{equation}
Up to main order, the off-diagonal block of the equation reads 
\begin{equation}
0=H_{a}w+wH_{b}+H_{s},\label{eq:linear}
\end{equation}
where we have defined $H_{b}\left(\boldsymbol{k}\right)\equiv H_{a}^{\ast}\left(-\boldsymbol{k}\right)$
and omitted $\boldsymbol{k}$ in the expression. And the diagonal
block of the expansion formula can be simplified as
\begin{equation}
h_{a}=H_{a}+\frac{1}{2}\left(H_{s}w^{\dagger}+wH_{s}^{\dagger}\right),\label{eq:pert}
\end{equation}
with the help of Eq.~(\ref{eq:linear}). It implies that the perturbative
squeezing interaction acts as an effective SOC $H_{\text{SOC}}\left(\boldsymbol{k}\right)\equiv\left(H_{s}w^{\dagger}+wH_{s}^{\dagger}\right)/2$,
which is added to the zeroth-order dispersion term $H_{a}\left(\boldsymbol{k}\right)$
in the effective Hamiltonian $h_{a}=H_{a}+H_{\text{SOC}}$,
since $H_{{\rm SOC}}\left(\bm{k}\right)$ involves the (quasi-)momentum
$\bm{k}$ as well as internal degrees of freedom including sublattices,
atomic orbitals and even spins. After some algebra, the explicit
expression of effective Hamiltonian in the $H_{a,b}$-representation
is then shown up, 
\begin{eqnarray}
\left(g_{a}^{\dagger}h_{a}g_{a}\right)_{ij} & = & D_{ai}\delta_{ij}-\frac{1}{2}\left[g_{a}^{\dagger}H_{s}\left(D_{ai}I+H_{b}\right)^{-1}H_{s}^{\dagger}g_{a}\right]_{ij}\nonumber \\
 &  & -\frac{1}{2}\left[g_{a}^{\dagger}H_{s}\left(D_{aj}I+H_{b}\right)^{-1}H_{s}^{\dagger}g_{a}\right]_{ij},\label{eq:expr}
\end{eqnarray}
where $D_{a,b}=g_{a,b}^{\dagger}H_{a,b}g_{a,b}$ are the diagonal
matrices of eigenenergies of $H_{a,b}$, $g_{a,b}$
are the unitary matrices and the subscripts $i,j$ denote the band
indices. In Eq. (\ref{eq:expr}), we have substituted
the unique solution, $\left(g_{a}^{\dagger}wg_{b}\right)_{ij}=-\left(H_{s}\right)_{ij}/\left(D_{ai}+D_{bj}\right)$.
Since $w$ is unique, the paraunitarity of $e^{W}$ is automatically
guaranteed. The squeezing-induced SOC described by the last two
terms records the virtual processes of bi-particle creation/annihilation
and annihilation/creation which are provoked by the squeezing interaction
$\frac{1}{2}\hat{\boldsymbol{a}}_{\boldsymbol{k}}^{\dagger}H_{s}\left(\boldsymbol{k}\right)\hat{\boldsymbol{a}}_{-\boldsymbol{k}}^{\dagger\top}+\text{h.c.}$.

The formula of Eq.~(\ref{eq:expr}), however, is too complex to grasp
its profoundness. It is expected to simplify via a slight
deformation with the topological feature unchanged. We first consider
a boring case that the dispersion term $H_{a}\left(\boldsymbol{k}\right)$
already has a finite band gap on high-lying excitation. In this case,
infinitesimal squeezing term$H_{s}\left(\boldsymbol{k}\right)$ can
be arbitrarily modified or even neglected. In the following, we focus
on another novel case that a band-crossing point of $H_{a}\left(\boldsymbol{k}\right)$
(labeled by $\left(\boldsymbol{p},\lambda\right)$ in the momentum-energy
space) is split by the squeezing interaction $H_{s}\left(\boldsymbol{k}\right)$,
with a bulk gap opened in the BE bands. This is similar to the band-inversion
phenomenon in fermionic topological insulators \citep{Kane2005PhysRevLett,Kane2005PhysRevLett2,Bernevig2006PRL,Bernevig2006Science,Fu2007PhysRevB}
and has covered most existed examples of topological bosonic BEs \citep{Di_Liberto2016PRL,Peano2016NatCommun,Xu2016PhysRevLett,Luo2018PRA}.
For simplicity, we merely discuss the case that the gap is opened
from a single excitation band-crossing point.

 Our strategy of simplification is to precisely characterize
the BE bands around the gap-opening point $\left(\boldsymbol{p},\lambda\right)$
but slightly deform the band structure elsewhere. This treatment grasps
the key topological origin that is the squeezing-induced gap opening.
According to the perturbation theory, the first-order correction toward
the zeroth-order degenerate point $\left(\boldsymbol{p},\lambda\right)$
is determined by block matrix 
\begin{equation}
\left(g_{a}^{\dagger}h_{a}g_{a}\right)_{ij}=D_{ai}\delta_{ij}-\left[g_{a}^{\dagger}H_{s}\left(\lambda I+H_{b}\right)^{-1}H_{s}^{\dagger}g_{a}\right]_{ij},\label{eq:block}
\end{equation}
where $i$ and $j$ are confined on the degenerate levels, i.e., $D_{ai}\left(\boldsymbol{p}\right)=D_{aj}\left(\boldsymbol{p}\right)=\lambda$.
Up to the main order, the energy splitting of $\left(\boldsymbol{p},\lambda\right)$
is accurately characterized by this formula. As for the points away
from $\left(\boldsymbol{p},\lambda\right)$, the correction from squeezing
is tiny and can be neglected or deformed in a topological sense. Therefore,
it is feasible to extend Eq.~(\ref{eq:block}) to the whole bands
($i,j$ running throughout the band indices), arriving at an approximate
formula 
\begin{equation}
h_{a}\left(\boldsymbol{k}\right)=H_{a}\left(\boldsymbol{k}\right)-\left[H_{s}\left(\lambda I+H_{b}\right)^{-1}H_{s}^{\dagger}\right]_{\boldsymbol{k}=\boldsymbol{p}}.\label{eq:approx}
\end{equation}
Note that the squeezing-induced correction $H_{s}\left(\lambda I+H_{b}\right)^{-1}H_{s}^{\dagger}$
becomes $\boldsymbol{k}$-independent. It thus can be interpreted
as an effective Zeeman coupling. Equation (\ref{eq:approx}) is the
key formula to understand the squeezing-induced topological gap on
bosonic BEs. Its applications and validity will be shown in two examples
below.

\subsection{Example of photonic crystal}

\begin{figure}
\includegraphics[width=1\columnwidth]{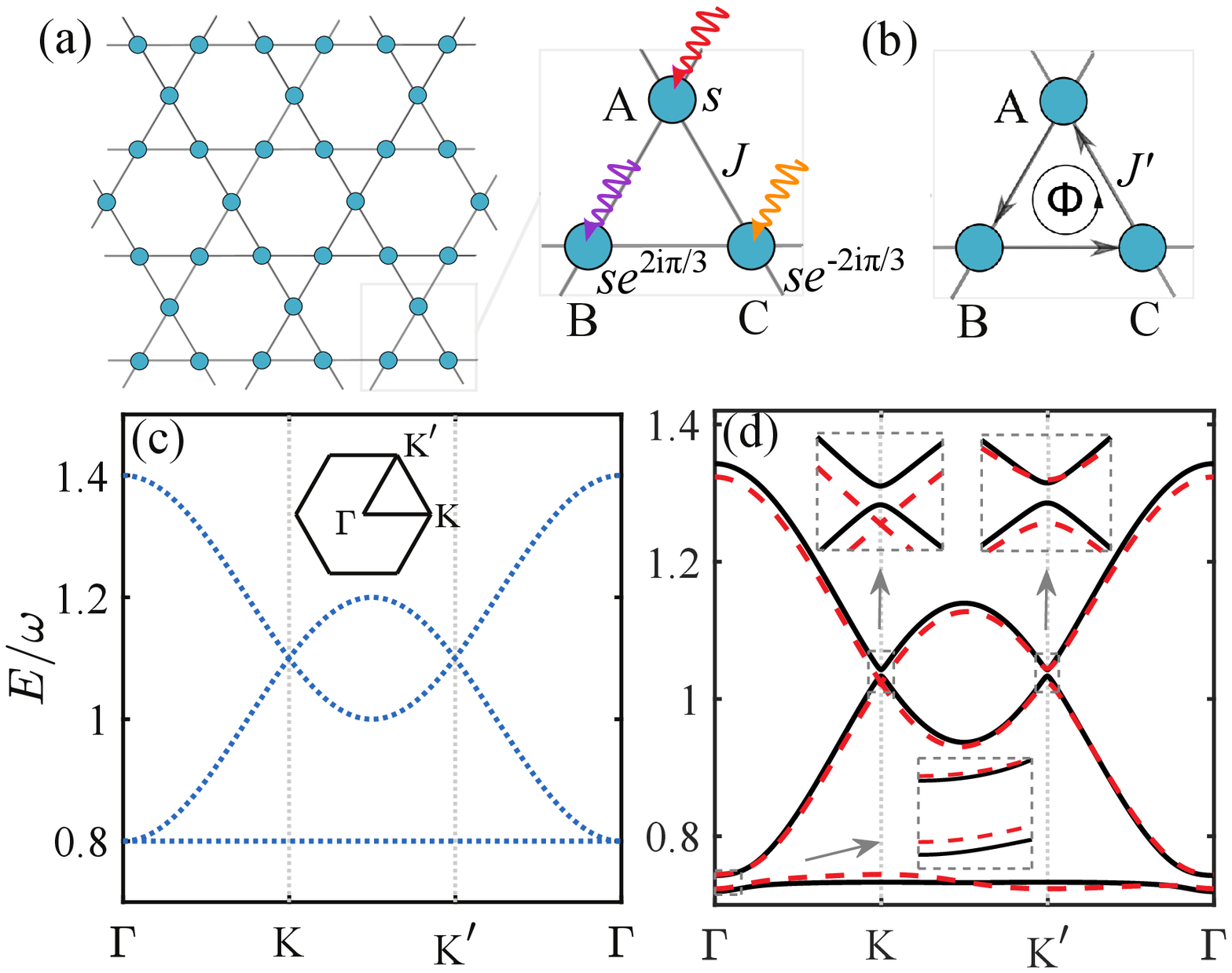}\caption{(color online) (a) Sketch of photonic kagome lattice with squeezing
interaction. The panel shows the squeezing interaction pattern and
hopping coupling. (b) A triangular plaquette of effective model (\ref{eq:OMN})
with hopping $J^{\prime}$, which has a nontrivial flux $\Phi$. Band
structures of (c) $H_{a}\left(\boldsymbol{k}\right)$ and (d) $h_{a}\left(\boldsymbol{k}\right)$
for the photonic crystal. In (c), the hexagon denotes the first Brillouin
zone. In (d), the black solid and red dashed lines denote the exact
and approximate BE spectra, respectively. Here, $J=0.1\omega$ and
$s=0.35\omega$.\label{fig:kag}}
\end{figure}

Firstly, we revisit topological BEs in photonic crystal with a kagome
lattice structure \citep{Peano2016NatCommun}, as shown in Fig.~\ref{fig:kag}(a).
It starts with a dielectric material containing periodic pores that
cause photonic band structure \citep{Joannopoulos2011}. In the presence
of crystalline defects in an appropriate pattern, localized modes
would appear and a tight-binding model could be utilized to describe
the system. Meanwhile, the nonlinear effect from $\chi^{\left(2\right)}$
optical medium would induce an onsite squeezing interaction with the
help of an auxiliary driving light \citep{Scully1997}. Via the rotating-wave
approximation of the driving field, the total Hamiltonian is given
by a bosonic BdG Hamiltonian. The dispersion and squeezing terms are
given by
\begin{eqnarray}
H_{a}\left(\bm{k}\right) & = & J\left(\begin{array}{ccc}
\omega/J & 1+e^{-\mathrm{i}\bm{k}\cdot\boldsymbol{l}_{3}} & 1+e^{\mathrm{i}\bm{k}\cdot\boldsymbol{l}_{2}}\\
 & \omega/J & 1+e^{-\mathrm{i}\bm{k}\cdot\boldsymbol{l}_{1}}\\
\text{h.c.} &  & \omega/J
\end{array}\right),\\
H_{s} & \equiv & s\left(\begin{array}{ccc}
1\\
 & e^{\mathrm{i}2\pi/3}\\
 &  & e^{-\mathrm{i}2\pi/3}
\end{array}\right),
\end{eqnarray}
where $\omega$ is the on-site energy, $J$ denotes the nearest-neighbor
hopping amplitude, and $s$ is the squeezing intensity. Here $\boldsymbol{k}=\left(k_{x},k_{y}\right)$
is the crystal momentum, and $\boldsymbol{l}_{1}=\left(1,0\right)$,
$\boldsymbol{l}_{2}=\left(-1/2,\sqrt{3}/2\right)$, $\boldsymbol{l}_{3}=\left(-1/2,-\sqrt{3}/2\right)$
are lattice vectors (for simplicity, the crystal constant is assumed
to be $1$.). The on-site squeezing interaction shows a phase distribution
in the internal space which is the key for generating the topological
BE.

The energy spectra of $H_{a}\left(\boldsymbol{k}\right)$ is gapless
as shown in Fig.~\ref{fig:kag}(c). There are two Dirac points at
$\left({\rm K},\lambda_{K}\right)$, $\left({\rm K^{\prime}},\lambda_{K}\right)$
and a quadratic band-crossing point at $\left(\Gamma,\lambda_{\Gamma}\right)$.
The BE spectra given by the eigenvalues of $h_{a}\left(\boldsymbol{k}\right)$
are numerically calculated via Eqs.~(\ref{eq:hamil}--\ref{eq:W}),
which are shown by the black solid lines in Fig.~\ref{fig:kag}(d).
We infer that the squeezing interaction splits all the degenerate
points and generates two bulk excitation gaps opening
at $\Gamma$, ${\rm K}$, ${\rm K}^{\prime}$ points. We focus on
the lowest excitation band and characterize the squeezing-induced
gap around $\left(\Gamma,\lambda_{\Gamma}\right)$ by Eq.~(\ref{eq:approx}).
After substituting $\left(\boldsymbol{p},\lambda\right)=\left(\Gamma,\lambda_{\Gamma}\right)=\left(\boldsymbol{0},\omega-2J\right)$
into the formula, we find that the effective non-interacting Hamiltonian
is described by 
\begin{equation}
h_{a}\left(\boldsymbol{k}\right)=H_{a}\left(\boldsymbol{k}\right)+J\left(\begin{array}{ccc}
-\epsilon/J & z & z^{\ast}\\
z^{\ast} & -\epsilon/J & z\\
z & z^{\ast} & -\epsilon/J
\end{array}\right),\label{eq:phot}
\end{equation}
with $\epsilon=\left|s\right|^{2}\omega/2\left(\omega^{2}-J\omega-2J^{2}\right)$
and $z=\left(\epsilon/\omega\right)\exp\left(-\mathrm{i}2\pi/3\right)$.
Its spectra are denoted by the red dashed line in Fig.~\ref{fig:kag}
(d). We find that the approximate result fits the exact solution well,
especially near the original band-crossing point $\left(\Gamma,\lambda_{\Gamma}\right)$.
The Chern number for lowest BE band is $-1$.  In addition,
it is worthy to mention that the analytical result disagrees at ${\rm K}$
point since the perturbation applied around $\Gamma$ point works for
the low-lying BEs and is not responsible for the high-lying BEs.

The value of Chern number can be analytically understood as follows.
At the symmetry point $\Gamma$, the effective Hamiltonian $h_{a}\left(\bm{k}=\Gamma\right)$
shows that each nearest-neighbor hopping given by $J\left(1+z/2\right)$
possesses a phase $\arg\left(1+z/2\right)$ and an flux $\Phi=3\arg\left(1+z/2\right)$
is accumulated for each triangular plaquette. This can be extended
from the symmetry point $\Gamma$ to the whole Brillouin zone in the
sense of topology. Therefore, we achieve a new model, as shown in
Fig.~\ref{fig:kag}(b), by making a slight deformation $z\mapsto z\left(1+e^{-\mathrm{i}\bm{k}\cdot\boldsymbol{l}_{i}}\right)/2$
for the squeezing-induced SOC term in Eq.~(\ref{eq:phot}). And it
is governed by the Hamiltonian 
\begin{equation}
h_{a}^{\prime}\left(\boldsymbol{k}\right)=\left(\begin{array}{ccc}
\omega-\epsilon & J^{\prime}\left(1+e^{-\mathrm{i}\bm{k}\cdot\boldsymbol{l}_{3}}\right) & J^{\prime\ast}\left(1+e^{\mathrm{i}\bm{k}\cdot\boldsymbol{l}_{2}}\right)\\
 & \omega-\epsilon & J^{\prime}\left(1+e^{-\mathrm{i}\bm{k}\cdot\boldsymbol{l}_{1}}\right)\\
\text{h.c.} &  & \omega-\epsilon
\end{array}\right),\label{eq:OMN}
\end{equation}
where the nearest-neighbor hopping amplitude is given by $J^{\prime}=J\left(1+z/2\right)$.
This would not affect the squeezing-induced topological gap at $\boldsymbol{k}=\boldsymbol{0}$
($\Gamma$ point) such that $h_{a}\left(\boldsymbol{k}\right)$ and
$h_{a}^{\prime}\left(\boldsymbol{k}\right)$ have the same topological
number. We notice that $h_{a}^{\prime}\left(\boldsymbol{k}\right)$
corresponds to an Ohgushi-Murakami-Nagaosa model \citep{Ohgushi2000PRB}
which has been well studied. It has a nonvanishing flux per triangular
plaquette given by $\Phi=3\arg\left(1+e^{-\mathrm{i}2\pi/3}\epsilon/2\omega\right)$
which breaks the time-reversal ($T$) symmetry, while the net flux
is zero. Its Chern number is then given by $C=\text{sgn}\left(\sin\Phi\right)=-1$
\citep{Ohgushi2000PRB} in the weak squeezing regime. Our analysis
also confirms the result that the squeezing interaction can be reduced
to an effective hopping \citep{Bardyn2016}.

\subsection{Example of ultracold atoms in optical lattice}

\begin{figure}
\includegraphics[width=1\columnwidth]{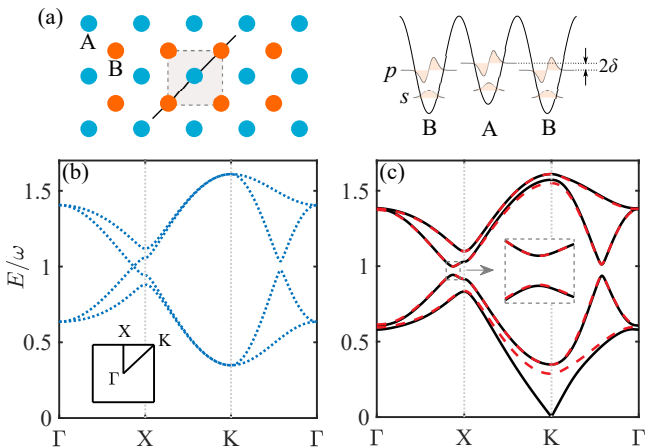}

\caption{(color online) (a) Sketch of the optical square lattice loaded cold
atoms \citep{Luo2018PRA}. The dashed square denotes the unit cell.
The right panel shows the optical potential along the direction of
the black solid line. Band structure of (b) $H_{a}\left(\boldsymbol{k}\right)$
and (c) $h_{a}\left(\boldsymbol{k}\right)$. In (b), the square denotes
the Brillouin zone. In (c), the black solid and red dashed lines denote
the exact and approximate BE spectra, respectively. The parameters
are given by $J_{1}=0.0555E_{R}$, $J_{2}=0.0912E_{R}$,
$J_{3}=0.0158E_{R}$, $J_{4}=-0.0098E_{R}$, $\delta=0.06E_{R}$,
$\omega=0.5789E_{R}$, $s_{1}=0.1E_{R}$, $s_{2}=-0.0106E_{R}$ and
$E_{R}$ is the recoil energy of $^{87}{\rm Rb}$ atom, which are the
same as Ref. \citep{Luo2018PRA}. \label{fig:sq}}
\end{figure}

Next, we revisit another model where bosonic atoms are loaded into
the second band of a square optical lattice \citep{Luo2018PRA}, as
shown in Fig.~\ref{fig:sq}(a). Focusing on the $p$-orbitals, the
single-particle Hamiltonian is then given by 
\begin{eqnarray}
H_{0}\left(\boldsymbol{k}\right) & = & \left(J_{3}+J_{4}\right)\left(\cos k_{x}+\cos k_{y}\right)I_{4}+\delta\sigma_{z}\otimes I_{2}\nonumber \\
 &  & +\left(J_{3}-J_{4}\right)\left(\cos k_{x}-\cos k_{y}\right)I_{2}\otimes\sigma_{z}\nonumber \\
 &  & +4J_{1}\cos\frac{k_{x}}{2}\cos\frac{k_{y}}{2}\sigma_{x}\otimes I_{2}\nonumber \\
 &  & -4J_{2}\sin\frac{k_{x}}{2}\sin\frac{k_{y}}{2}\sigma_{x}\otimes\sigma_{x}.
\end{eqnarray}
Here $\sigma_{x,z}$ are the conventional Pauli matrices. $2\delta$
is the offset energy of the double-well. $J_{1}$ and $J_{2}$ are
the amplitudes of the nearest-neighbor hopping, while $J_{3}$ and
$J_{4}$ are the ones of next-nearest-neighbor hopping. (We set the
length of lattice vector as $1$ for simplicity.) The interaction
among atoms generates a spontaneous time-reversal symmetry breaking,
leading to a chiral bosonic superfluid \citep{Volovik2003}. Then
after a mean-field approximation, the BEs on top of the chiral superfluid
are described by a bosonic BdG Hamiltonian. As a difference from the
above photonic crystal model, the complete gapless dispersion term
$H_{a}\left(\boldsymbol{k}\right)$ here contains the tight-binding
Hamiltonian $H_{0}\left(\bm{k}\right)$ and also the contribution
from the contact atomic interaction. The dispersion and squeezing
terms of the BdG Hamiltonian are given by
\begin{eqnarray}
H_{a}\left(\boldsymbol{k}\right) & = & H_{0}\left(\bm{k}\right)+\omega I_{4}+\Delta\sigma_{z}\otimes I_{2}\nonumber \\
 & = & \left[\omega+\left(J_{3}+J_{4}\right)\left(\cos k_{x}+\cos k_{y}\right)\right]I_{4}\nonumber \\
 &  & +\left(\Delta+\delta\right)\sigma_{z}\otimes I_{2}\nonumber \\
 &  & +\left(J_{3}-J_{4}\right)\left(\cos k_{x}-\cos k_{y}\right)I_{2}\otimes\sigma_{z}\nonumber \\
 &  & +4J_{1}\cos\frac{k_{x}}{2}\cos\frac{k_{y}}{2}\sigma_{x}\otimes I_{2}\nonumber \\
 &  & -4J_{2}\sin\frac{k_{x}}{2}\sin\frac{k_{y}}{2}\sigma_{x}\otimes\sigma_{x},
\end{eqnarray}
\begin{equation}
H_{s}\equiv-s_{1}\sigma_{z}\otimes\sigma_{z}+\mathrm{i}s_{1}I_{2}\otimes\sigma_{x}-s_{2}I_{2}\otimes\sigma_{z}+\mathrm{i}s_{2}\sigma_{z}\otimes\sigma_{x},
\end{equation}
where $s_{1,2}$ are the interaction intensities and $\Delta=4s_{2}$,
and $\omega$ is the parameter depending on both the tight-binding
model and atomic interaction.

The energy spectra of $H_{a}\left(\boldsymbol{k}\right)$ and $h_{a}\left(\boldsymbol{k}\right)$
are presented in Fig.~\ref{fig:sq}(a--b), respectively. We see
that $H_{a}\left(\boldsymbol{k}\right)$ is gapless, while $h_{a}\left(\boldsymbol{k}\right)$
has a high-lying excitation gap opened by the squeezing interaction.
We then calculate the approximate BE spectra based on Eq.~(\ref{eq:approx})
and present the result by the red dashed lines shown in Fig.~\ref{fig:sq}(b).
It coincides with the exact solution (black solid lines) in most area,
except the vicinity of zero energy at ${\rm K}$ point ($\bm{k}=\left(\pi,\pi\right)$).
Whereas, the Goldstone mode far away from the gap is unimportant for
our topological analysis. The Chern number of the excitation bands
below the gap is $-2$.

Interestingly, in the single-particle picture, the band structure
of $H_{0}\left(\bm{k}\right)$ is gapped \citep{Luo2018PRA}, while
the contribution of interaction to $H_{a}\left(\bm{k}\right)$ results
in the dispersion gapless, and the presence of the squeezing interaction
$H_{s}$ opens a high-lying topological excitation gap. In other words,
the interaction-driven topological model must experience the process
of gap closing-and-opening. This topological phase transition can
be understood in the framework of AZ class \citep{Kitaev2009AIP,Stone2011JPA,Chiu2016RevModPhys}
since the the effective model $h_{a}\left(\bm{k}\right)$ is non-interacting.
In $h_{a}\left(\bm{k}\right)$, the term $\Delta\sigma_{z}\otimes I_{2}$
contributed from interaction closes the band gap shown in Fig. \ref{fig:sq}(b).
And then the squeezing-induced Zeeman coupling breaks the time-reversal
symmetry and opens an excitation gap at $\left(\bm{p},\lambda\right)$
point shown in Fig.~\ref{fig:sq}(c). It acts as the role of band-inversion
\citep{Kane2005PhysRevLett,Kane2005PhysRevLett2,Bernevig2006PRL,Fu2007PRL}
and results in the nonvanishing Chern number.

Till now, the validity and power of Eq.~(\ref{eq:approx}) have been
verified. It lays a foundation for analyzing and constructing topological
bosonic BEs.

\section{Minimal constructions\label{sec:Constr}}

In this section, we search for the minimal construction of both $\mathbb{Z}$-type
and $\mathbb{Z}_{2}$-type topological bosonic BEs. The minimal model
is expected to be analytically solvable with an intuitive physical
picture. This is beneficial to exploring more complicated model in
realistic experiments. Like the simplest free-fermion models \citep{Stone2011JPA},
we adopt the elegant Clifford algebra to build a two-band bosonic
BdG Hamiltonian. Meanwhile, a momentum-independent squeezing interaction
is designed for generating topological BEs. To be more specific, $H_{a}\left(\boldsymbol{k}\right)$
is assumed to be 
\begin{equation}
H_{a}\left(\bm{k}\right)=\lambda I+\boldsymbol{r}\left(\boldsymbol{k}\right)\cdot\boldsymbol{\gamma},\label{eq:disper}
\end{equation}
where $\boldsymbol{r}\left(\boldsymbol{k}\right)=\left(r_{1},r_{2},\cdots\right)$
is a set of real functions of $\boldsymbol{k}$ and $\boldsymbol{\gamma}=\left(\gamma_{1},\gamma_{2},\cdots\right)$
is a set of Clifford generators satisfying $\left\{ \gamma_{m},\gamma_{n}\right\} =2\delta_{mn}I$.
The domain of $\boldsymbol{k}$ can either be a compactized Euclidean
space $\mathbb{R}^{d}\cup\infty=S^{d}$ (for free-propagating bosons)
or reciprocal space $T^{d}$ (for lattice models). We further assume
one band-crossing point determined by $\boldsymbol{r}\left(\boldsymbol{p}\right)=\boldsymbol{0}$
which locates at a inversion-invariant (crystal) momentum $\boldsymbol{p}=-\boldsymbol{p}$.
The squeezing interaction is chosen as the simplest on-site interaction
\begin{equation}
H_{s}\equiv\xi I+\boldsymbol{s}\cdot\boldsymbol{\gamma},\label{eq:squeez}
\end{equation}
where $\xi$ is a complex number and $\boldsymbol{s}$ is a real vector
satisfying $(\boldsymbol{s}\cdot\boldsymbol{\gamma})^{\top}=\boldsymbol{s}\cdot\boldsymbol{\gamma}$.
The transpose symmetry $H_{s}=H_{s}^{\top}$ guarantees that $H\left(\boldsymbol{k}\right)$
obeys the particle-hole symmetry described by Eq.~(\ref{eq:symm}).

Applying the perturbation theory discussed in Sec.~\ref{sec:Pertur},
we obtain the effective non-interacting Hamiltonian which is given
by 
\begin{equation}
h_{a}\left(\bm{k}\right)=\eta I+\boldsymbol{r}\left(\boldsymbol{k}\right)\cdot\boldsymbol{\gamma}-\frac{\text{Re}\xi}{\lambda}\boldsymbol{s}\cdot\boldsymbol{\gamma},\label{eq:Dirac}
\end{equation}
where $\eta=\lambda-(\left|\xi\right|^{2}+\left\Vert \boldsymbol{s}\right\Vert ^{2})/2\lambda$.
The squeezing-induced correction simply reduces to a mass term $-\text{Re}\xi\left(\boldsymbol{s}\cdot\boldsymbol{\gamma}\right)/\lambda$,
which opens a topological excitation gap around the original band-crossing
point, e.g. Dirac point. We immediately infer that the effective Hamiltonian
of bosons reproduces the minimal construction of fermionic topological
insulators. We can utilize the well-studied fermionic models to create
the related bosonic versions for generating topological BEs, and two
examples are given below.

\subsection{$\mathbb{Z}$-type topological BE}

\begin{figure}
\includegraphics[width=1\columnwidth]{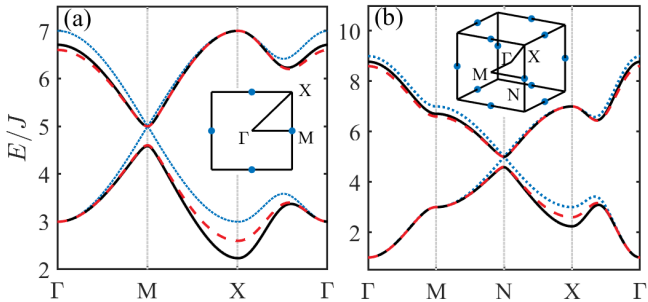}\caption{(color online) Band structures of (a) $\mathbb{Z}$-type and (b) $\mathbb{Z}_{2}$-type
topological bosonic BE. The blue dotted lines denote the spectra of
$H_{a}\left(\boldsymbol{k}\right)$, while the black solid and red
dashed lines denote the exact and approximate spectra of $h_{a}\left(\boldsymbol{k}\right)$,
respectively. The approximate result is from Eq. (\ref{eq:Dirac}).
The square and cube denote the Brillouin zones in which the blue dots
denote the locations of Dirac points. Parameters are $\lambda=5J$
and $\xi=s=J$.\label{fig:constr}}
\end{figure}

Firstly, we attempt to construct a two-dimensional bosonic BdG model
which supports a $\mathbb{Z}$-type topological BE. The Hamiltonian
has no other symmetry except the intrinsic particle-hole symmetry
and the topological invariant of $h_{a}\left(\boldsymbol{k}\right)$
is characterized by a Chern number. The Hamiltonian in form of Eqs.~(\ref{eq:disper}--\ref{eq:squeez})
is set as \citep{Qi2006PhysRevB}
\begin{equation}
\bm{r}\left(\bm{k}\right)=J\left(\sin k_{x},\sin k_{y},\cos k_{x}+\cos k_{y}\right),
\end{equation}
 and $\boldsymbol{s}=\left(0,0,s\right)$. The Clifford generators
are chosen as Pauli matrices, i.e., $\boldsymbol{\gamma}=\left(\sigma_{1},\sigma_{2},\sigma_{3}\right)$.
The spectra of $H_{a}\left(\boldsymbol{k}\right)$ and $h_{a}\left(\boldsymbol{k}\right)$
are compared in Fig.~\ref{fig:constr}(a).

We obtain two Dirac points of $H_{a}\left(\boldsymbol{k}\right)$
from $\boldsymbol{r}\left(\boldsymbol{k}\right)=\boldsymbol{0}$ which
locate at $\left(0,\pi\right)$ and $\left(\pi,0\right)$. In the
presence of squeezing, the mass term $-\left(s\text{Re}\xi/\lambda\right)\sigma_{3}$
derived from Eq.~(\ref{eq:Dirac}) may open an excitation gap and
extremely changes the Berry curvature close to each Dirac cone. The
gapped condition of the effective Hamiltonian $h_{a}\left(\boldsymbol{k}\right)$
can be calculated as $\left|s\text{Re}\xi/\lambda J\right|\neq2$.
The Chern number of the lower excitation band is given by 
\begin{equation}
C=\begin{cases}
-\text{sgn}\left(s\text{Re}\xi/\lambda J\right), & \left|s\text{Re}\xi/\lambda J\right|<2,\\
0, & \left|s\text{Re}\xi/\lambda J\right|>2.
\end{cases}
\end{equation}
It equals to the winding number (Brouwer degree \citep{Brouwer1912})
of $\boldsymbol{r}\left(\boldsymbol{k}\right)-\frac{\text{Re}\xi}{\lambda}\boldsymbol{s}$
which maps $T^{2}$ to $\mathbb{R}^{3}\backslash\boldsymbol{0}\approx S^{2}$.
This model shows that the simplest Dirac-like construction is able
to generate nontrivial bosonic BE.

\subsection{$\mathbb{Z}_{2}$-type topological Bogoliubov excitations}

Next, we construct a three-dimensional model with time-reversal symmetry,
whose topology is characterized by a $\mathbb{Z}_{2}$ invariant.
Like the above construction, the Hamiltonian in form of Eqs.~(\ref{eq:disper}--\ref{eq:squeez})
is chosen as \citep{Bernevig2006PRL}
\begin{equation}
\bm{r}\left(\bm{k}\right)=J\left(\sin k_{x},\sin k_{y},\sin k_{z},\sum_{n=x,y,z}\cos k_{n}+1\right),
\end{equation}
and $\boldsymbol{s}=\left(0,0,0,s\right)$. The parameter $\xi$ is
set as a real number. The Clifford generators $\bm{\gamma}=\left(\gamma_{1},\gamma_{2},\gamma_{3},\gamma_{4}\right)$
are 
\begin{equation}
\gamma_{4}=\left(\begin{array}{cc}
I_{2}\\
 & -I_{2}
\end{array}\right),\quad\gamma_{n}=\left(\begin{array}{cc}
 & \sigma_{n}\\
\sigma_{n}
\end{array}\right),
\end{equation}
where $n=1,2,3$. The Hamiltonian $H\left(\boldsymbol{k}\right)$
not only satisfies the particle-hole symmetry Eq.~(\ref{eq:symm})
but also respects $T$ symmetry $U_{T}^{-1}H^{\ast}\left(-\boldsymbol{k}\right)U_{T}=H\left(\boldsymbol{k}\right)$
with $U_{T}=I_{4}\otimes\sigma_{2}$. We also confirm that the effective
Hamiltonian $h_{a}\left(\boldsymbol{k}\right)$ given by Eq.~(\ref{eq:Dirac})
preserves the $T$ symmetry, i.e., $u_{T}^{-1}h_{a}^{\ast}\left(-\boldsymbol{k}\right)u_{T}=h_{a}\left(\boldsymbol{k}\right)$
with $u_{T}=I_{2}\otimes\sigma_{2}$.

Figure~\ref{fig:constr}(b) illustrates the spectra of $H_{a}\left(\boldsymbol{k}\right)$
and $h_{a}\left(\boldsymbol{k}\right)$. There are three Dirac points
located at $\left(0,\pi,\pi\right)$, $\left(\pi,0,\pi\right)$ and
$\left(\pi,\pi,0\right)$. The squeezing-induced mass term opens a
gap close to these Dirac cones if $\left|\xi s/\lambda J-1\right|\neq1,3$.
The effective Hamiltonian $h_{a}\left(\boldsymbol{k}\right)$ thus
describes a $T$-invariant topological insulator. Its energy bands
have Kramer's degeneracy so that the Chern number of the lowest bands
always vanishes. The corresponding topological feature is characterized
by a $\mathbb{Z}_{2}$ invariant which is given by 
\begin{equation}
\nu=\begin{cases}
1, & \left|\xi s/\lambda J-1\right|\in\left(1,3\right),\\
0, & \left|\xi s/\lambda J-1\right|\in\left[0,1\right)\cup\left(3,+\infty\right).
\end{cases}\label{eq:Z2}
\end{equation}
This model shows that the minimal construction is able to generate
various types of topological bosonic BEs.

Our construction is potentially realizable in bosonic systems. The
platforms like optical lattices with cold atoms \citep{Jaksch1998PhysRevLett,Yukalov2009LaserPhys,Zhang2018AdvPhys},
magnonic crystals \citep{Matsumoto2011PRL,Gulyaev2003JETPL,Singh2004NanoTech,Wang2006NanoTech,Lenk2011PhyRep,Adeyeye2008JPD}
and photonic crystals \citep{Wang2009Nature,Joannopoulos2011,Wang2013PhysRevB,Notomi2008NatPhotonics,Eggleton2011NatPhotonics}
are accessible for the single-particle models. Whereas, the actual
challenge for experiments is the realization of the required squeezing
interaction. Apart from magnonic materials in which topological gap
and Weyl points have been observed with the aid of theoretical calculations
\citep{Chisnell2015PRL,McClarty2017NatPhys,Bao2018NatComm,Yao2018NatPhys,Chen2018PRX},
here we predict that the photonic crystals might be another candidate.
The photonic array is made of optically nonlinear materials with second-order
nonlinear susceptibility $\chi^{\left(2\right)}$. When the system
is coherently pumped with the frequency matching condition (i.e.,
one photon resonantly converting to two photons), the auxiliary mode
can be treated classically and the squeezing interaction then emerges
\citep{Scully1997}. Such a squeezing can be tuned by adjusting the
amplitude and phase of the pump. And the construction based on $\bm{k}$-independent
squeezing interaction would make the implementation easier.

\section{Conclusions\label{sec:Conclu}}

In summary, we have established a framework for the topological analysis
and construction of bosonic BdG models in the weak interaction limit.
An effective non-interacting Hamiltonian has been derived from a squeezing
transformation. After a perturbation truncation, the squeezing interaction
has been found to play an essential role as an effective SOC. For
the squeezing-induced topological gap opening upon a gapless band
dispersion, the effective SOC has further been simplified as a momentum-independent
Zeeman coupling. Then the formulation has been applied to two existed
models, which gives deeper understandings on their topological structures.
Finally, the minimal construction of various types of topological
bosonic BEs has also been built based on the Clifford algebra and
two concrete examples have been provided, which are applicable for
experiments.

Our work opens up the studies of squeezed topological phases in future.
Firstly, the squeezing properties of the bulk and edge BE modes could
be further investigated based on the squeezing transformation. Since
the BdG system can be transformed to a non-interacting one, the bulk-edge
correspondence for bosonic system may be strictly established. Secondly,
the $\mathbb{Z}_{2}$-type topological bosonic BE has very few examples
at the present stage. It is probable to find more examples in photonic
crystal and cold atom system with the help of our perturbation formula.
Lastly, the topological analysis toward the gapless points in bosonic
BE bands is also an interesting topic.
\begin{acknowledgments}
We thank Guang-Quan Luo for helpful discussions and providing detail
parameters on the ultracold atom models. This work is supported by
National Key R\&D Program of China (Grant No.~2018YFA0307200), the
Key-Area Research and Development Program of GuangDong Province (Grant
No.~2019B030330001), National Natural Science Foundation of China
(Grant No.~U1801661), and high-level special funds from Southern
University of Science and Technology (Grant No.~G02206401).
\end{acknowledgments}

\appendix

\section{Generic quadratic bosonic model\label{sec:Gen}}

Here we discuss the generic case that matrix $H\left(\boldsymbol{k}\right)=\left(\begin{array}{cc}
H_{a} & H_{s}\\
H_{s}^{\dagger} & H_{b}
\end{array}\right)$ has no symmetric constraint. Now the field operator turns to $\hat{\phi}_{\bm{k}}^{\dagger}=\left(\hat{\bm{a}}_{\bm{k}}^{\dagger},\hat{\bm{b}}_{-\bm{k}}\right)$
which contains independent creation part $\hat{\bm{a}}_{\bm{k}}^{\dagger}$
and annihilation part $\hat{\bm{b}}_{\bm{k}}=\left(\hat{b}_{\bm{k},1},\cdots,\hat{b}_{\bm{k},N^{\prime}}\right)$.
And the Hamiltonian is written as 
\begin{equation}
\hat{H}=\sum_{\bm{k}}\hat{\phi}_{\bm{k}}^{\dagger}H\left(\bm{k}\right)\hat{\phi}_{\bm{k}}
\end{equation}
without factor $1/2$. Physically, this system consists of a pair
of bosonic tight-binding models $\sum_{\boldsymbol{k}}\hat{\boldsymbol{a}}_{\boldsymbol{k}}^{\dagger}H_{a}\left(\boldsymbol{k}\right)\hat{\boldsymbol{a}}_{\boldsymbol{k}}$
and $\sum_{\boldsymbol{k}}\hat{\boldsymbol{b}}_{-\boldsymbol{k}}H_{b}\left(\boldsymbol{k}\right)\hat{\boldsymbol{b}}_{-\boldsymbol{k}}^{\dagger}$
coupled by two-mode squeezing interactions $\sum_{\boldsymbol{k}}\hat{\boldsymbol{a}}_{\boldsymbol{k}}^{\dagger}H_{s}\left(\boldsymbol{k}\right)\hat{\boldsymbol{b}}_{-\boldsymbol{k}}^{\dagger}+\text{h.c.}$.
This construction in zero dimension has been widely studied in quantum
optics .

Similar to the above treatment, we adopt a new squeeze operator 
\begin{equation}
\hat{S}=\exp\sum_{\boldsymbol{k}}\left[\boldsymbol{a}_{\boldsymbol{k}}^{\dagger}w\left(\boldsymbol{k}\right)\boldsymbol{b}_{-\boldsymbol{k}}^{\dagger}-\text{h.c.}\right]
\end{equation}
to achieve the previous transformation 
\begin{equation}
\hat{S}^{\dagger}\hat{H}\hat{S}=\sum_{\boldsymbol{k}}\hat{\phi}_{\boldsymbol{k}}^{\dagger}h\left(\boldsymbol{k}\right)\hat{\phi}_{\boldsymbol{k}},\,h=e^{W}He^{W},
\end{equation}
and keep choosing $W$ by Eq. (\ref{eq:W}), arriving at block-diagonal
matrix $h\left(\boldsymbol{k}\right)=\left(\begin{array}{cc}
h_{a}\left(\bm{k}\right)\\
 & h_{b}\left(\bm{k}\right)
\end{array}\right)$. Eventually, the Hamiltonian in the squeezed state representation
consists of two isolated non-interacting systems, given by 
\begin{equation}
\hat{S}^{\dagger}\hat{H}\hat{S}=\sum_{\boldsymbol{k}}\hat{\boldsymbol{a}}_{\boldsymbol{k}}^{\dagger}h_{a}\left(\boldsymbol{k}\right)\boldsymbol{a}_{\boldsymbol{k}}+\sum_{\boldsymbol{k}}\hat{\boldsymbol{b}}_{-\boldsymbol{k}}h_{b}\left(\boldsymbol{k}\right)\boldsymbol{b}_{-\boldsymbol{k}}^{\dagger}.
\end{equation}

The topological information of quadratic bosonic system $\hat{H}$
is fully contained in the effective Hamiltonian $h\left(\boldsymbol{k}\right)=h_{a}\left(\boldsymbol{k}\right)\oplus h_{b}\left(\boldsymbol{k}\right)$,
characterized by two AZ classes. The approximate formula of $h_{a}\left(\boldsymbol{k}\right)$
is still given by Eq. (\ref{eq:approx}) if the system has the squeezing-induced
gap. And $h_{b}\left(\boldsymbol{k}\right)$ can be approximated by
the same procedure, i.e., 
\begin{equation}
h_{b}\left(\boldsymbol{k}\right)=H_{b}\left(\boldsymbol{k}\right)-\left[H_{s}^{\dagger}\left(H_{a}+\lambda^{\prime}I\right)^{-1}H_{s}\right]_{\boldsymbol{k}=\boldsymbol{p}^{\prime}},
\end{equation}
where $\left(\boldsymbol{p}^{\prime},\lambda^{\prime}\right)$ is
a band-crossing point of $H_{b}\left(\boldsymbol{k}\right)$.

\bibliographystyle{apsrev4-1}
\bibliography{Sqz}

\end{document}